\documentstyle[aps]{revtex}

\def \bi{\bibitem}
\def\e{{\rm e}}
\newcommand {\hh} { {\tt h}}

\begin{document}
\title{Quasi-equilibrium interpretation of aging dynamics}
\author{ S. Franz, M. A. Virasoro\\
}
\address{
The
Abdus Salam International Center for Theoretical Physics\\
Strada Costiera 11,
P.O. Box 563,
34100 Trieste (Italy)\\
}
\date{\today}
\maketitle

\begin{abstract}
We develop an interpretation of the off-equilibrium dynamical solution
of mean-field glassy models in terms of quasi-equilibrium concepts. We show
that the relaxation of the ``thermoremanent magnetization" follows
a generalized version of the Onsager regression postulate of
induced fluctuations. We then find the rationale for the equality
between the {\it
fluctuation-dissipation ratio} and the rate of growth
of the configurational entropy close to the asymptotic state,
found empirically in mean-field solutions.
\end{abstract}


\section*{}

\section{Introduction}

Aging is a scaling dynamical regime characteristic of glassy systems
\cite{aging}. In this regime, typical features of equilibrium systems, such as
the asymptotic absence of macroscopic heat currents,
coexist with non-stationary aspects such as the dependence of
the correlation and response functions on the system's ``age'',
i.e. the time spent in the low temperature phase.  The solution of
mean-field spin glass models \cite{cuku,frame,cukusk,francesi} has given a
general framework to understand aging phenomena, and has produced
detailed predictions, which have been verified in numerical
simulations of long-range \cite{felix} and short-range \cite{x(q)-sim}
glassy systems. A characteristic prediction of this solution is the
existence, at low temperature, of a dynamical regime where
extensive quantities depending on the configuration of the system at a
single time (one time observables in the following) are well
thermalized, (or evolve very slowly), while two time correlation function
and susceptibilities exhibit non-stationary scale invariant behavior.

Despite this coherent theoretical scheme, and recent progress in
linking aging dynamics to the nature of the equilibrium regime
\cite{fmpp}, a physical
understanding of some fundamental aspects of aging dynamics is still
lacking. While the investigation of both equilibrium and asymptotic
off-equilibrium regimes give solutions that show unexpected
coincidences, all efforts to interpret aging as a quasi-equilibrium
condition have been thwarted by facts such as:
\begin{enumerate}
\item
No matter how large we take the aging time if we then wait long enough the
system eventually wanders away from any finite region of phase
space \cite{cuku}.
\item
Two identical systems starting from the same condition at any given
aging time will always come apart as far as possible \cite{BBM}.
\end{enumerate}

In this paper we try to make ends meet. In particular, we try to give a
physical and intuitive explanation of the link between the so called
``fluctuation dissipation ratio'' (FDR) (the factor $x(q)$) and the
Parisi function (in Sherrington-Kirkpartick like models) or the derivative
of the configurational entropy close to the threshold state (in
p-spin-like  models). Our work should  be understood as a physical
interpretation of mean field aging dynamics in terms of a
quasi-equilibrium scenario and not as an alternative derivation of the
results of the theory.

We show that a modified version of the Onsager postulate on regression
of fluctuations applies to the relaxation of the
magnetization in thermoremanent magnetization experiments.  The
observed anomalies in the response are then analyzed. In
this paper we discuss aging mainly in the case of a ``one time
sector'' approximation, which is the dynamic counterpart
of the ``one step replica symmetry breaking'' (1RSB) approximation
in the equilibrium analysis. This is exact in models like
the p-spin model,  while it is only an approximation, and a rather crude
one, when
continuous replica symmetry breaking is present, like in the
Sherrington-Kirkpatrick (SK) model or the random manifold model.
 We will
refer to the first class of models as ``p-spin-like'' and to the
second as ``SK-like''.
We discuss only this ``one time sector'' approximation to unify the
argument, and
simplify the notation. But we expect the reader to be able to
generalize it without effort.

The picture that emerges from our analysis is simple and intuitive:
the age of an aging systems determines the rate of entropy decrease,
i.e. the flow rate of heat towards the thermal bath. A small force in
the linear response regime cannot change this rate. Acquiring a non
zero magnetization means entropy reduction which has then to be
compensated by an increase (or reduced decrease) of the free-energy
associated with the spin-couplings. As a consequence, the response
becomes proportional to the growth of the logarithm of the number of
{\it quasi-states} (to be defined later) with free-energy. It is as if
the slow degrees of freedom respond to external forces by sampling
states that lie above in free-energy while they are blocked from
exploring those that are at the same or at a lower level. This paper
presents what we believe are convincing arguments in favor of this
assertion. As a byproduct the time scale dependent effective
temperatures will appear \cite{teff} and their connection with the
derivative of the logarithm of the number of quasi-states (or
configurational entropy) with respect to free-energy is explained.
\cite{storia}

We organize the paper as follows: in section II we review some
properties of aging in mean-field. In section III we discuss our
definition of quasi-equilibrium and how it relates to dynamics. In
section IV we recall Onsager regression postulate and generalize it to
aging systems. In section V we discuss the origin of the FDR in
SK-like models, while we treat the case of p-spin-like models in
section VI.  Finally we present a summary and the conclusions.

\section{A short review}

We consider a mean-field spin-glass quenched at a given time $t=0$ into the
glassy phase. For simplicity we will
imagine that the degrees of freedom consist of Ising spins
$S_i=\pm 1$, $i=1,...,N$ ($N\to \infty$). We are interested on the
long time dynamics of such a system, i.e. on a regime where it has
thermalized for a long time $t_w$ before any measurements.  The long
time limit $t_w\to\infty$ is always taken after the
thermodynamic limit.

The free-energy, and its derivatives with respect
to the control parameters (e.g. temperature) tend to some
asymptotic time-independent values. More interesting is the behavior of
observables depending on two time
variables, such as correlation and response functions.  Aging behavior
manifests itself as an asymptotic non-stationary dependence on time of
these quantities.

Let us consider the spin-spin time dependent autocorrelation function
\begin{equation}
C(t,u)=\frac 1 N \sum_{i=1}^N S_i(t) S_i(u)\;\;\; t>u>>t_w
\end{equation}

A first, short time,  dynamical  regime is obtained considering the
difference $\tau=t-u$ finite to derive
stationary correlations $C_{st}^*(\tau)$. Let us denote  $q_{EA}$
the long $\tau$ limit of $C_{st}^*$.
In the  aging  regime  $C(t,u)$ relaxes below $q_{EA}$.
In the ``one time
sector'' approximation scheme,  the
correlation function in this regime can  be written as \cite{cuku},
\begin{equation}
C(t,u)=C_{ag}(\hh (u)/\hh (t))
\label{co}
\end{equation}
where $\hh (\cdot)$ is an increasing function not derivable from the
present theory, and $t,u$ are large with
$\lim_{t,u\to\infty}\hh (u)/\hh (t)=finite$. The formulae in the two regimes
are summarized in:
\begin{equation}
C(t,u)=C_{st}(t-u)+C_{ag}(\hh (u)/\hh (t))
\label{corr}
\end{equation}
where $C_{st}(t-u)=C^*_{st}(t-u)-q_{EA}$ is a monotonically decreasing
function equal to $1-q_{EA}$ for $t-u=0$ and tending to zero for
$t-u\to\infty$, while $C_{ag}(\hh (u)/\hh (t))$ is equal to $q_{EA}$ for
$\hh (u)/\hh (t)=1$ $(u=t)$ and tends to $q_0$ for $\hh (u)/\hh (t)\to 0$.
In order to simplify the notation we will suppose that
the ``time reparametrization'' $\hh (t)$ is the identity $\hh (t)=t$,
and $C_{ag}(\hh (u)/\hh (t))=C_{ag}(u/t)$. We will also suppose $q_0=0$, but
this will not affect any of our following arguments.

We stress that the form (\ref{corr}) of the correlation function
implies that if we fix the value of $u$ and let $t$ run, the time
spent at the value of the correlation equal to $q_{EA}$ is much larger
then the time needed to reach it. This is an aspect of the ``time
scale separation'' observed in glassy systems that will play a crucial role
in our discussion.

We are also interested in  the behavior of the linear response function,
\begin{equation}
R(t,u)=\frac 1 N \sum_{i=1}^N \frac{\delta}{\delta h_i(u)}
\langle S_i(t)\rangle|_{h=0} \;\;\; t>u>>t_w
\end{equation}
and the corresponding integrated function
\begin{equation}
\chi(t,u)=\int_{t_w}^u ds\; R(t,s) \;\;\; t>u>>t_w
\end{equation}
which represent the susceptibility  at time $t$ in a ``thermoremanent
magnetization'' experiment in which a constant small field has acted from
the time $t_w$ up to time $u$. By the condition $u>>t_w$ we mean
$t_w/u\to 0$ for $t_w\to \infty$.
The linear response theory requires that the limit $h\to 0$
be  taken {\it before} sending the time $u$ to infinity.

While in the stationary regime the fluctuation dissipation relation
$R(t,s)=R_{st}(t-s)=\beta \frac{\partial C(t,s)}{\partial s}$ is
verified, in the aging regime the relation is substituted by having
a non trivial fluctuation-dissipation ratio (FDR):
\begin{equation}
x(q)=\lim_{t,s\to \infty \atop C(t,s)=q} \frac{TR(t,s)}
{\frac{\partial C(t,s)}{\partial s}},
\end{equation}
which turns out to coincide with the function $x(q)$ that appears in the
replica approach that in principle applies to equilibrium of
different kinds \cite{BaVi,remi}.
In the one sector scenario $x(q)$ is constant all through the aging
regime.

The FDR $x(q)$ verifies the mathematical properties of a cumulative
probability distribution, a   feature that  has been
explained in recent work where it has been shown that there is a deep
connection between the dynamic properties during aging and the
property of ergodicity breaking in equilibrium \cite{fmpp}.  Using
only the hypothesis of equilibration of one time observables (OTO) and
the existence of a linear response regime for the correlation
functions (stochastic stability), it was proved that the
function $x(q)$ is related to the function $P(q)$
describing the
statistics of equilibrium pure states \cite{parisi} through the equation:
\begin{equation}
P(q)=\frac{dx(q)}{dq}.
\end{equation}

The theorem was originally formulated for finite dimensional systems,
where the OTOs are guaranteed to thermalize but can be generalized to
mean-field long range models of the SK-like class. A different situation is
found in p-spin-like models
\cite{pspin} where one of the
hypotheses of the theorem is violated because the
asymptotic value of the dynamic energy is higher than the one of the
states dominating the partition function.

\section{The quasi-equilibrium hypothesis}

From now on we will work on the ``one time sector'' approximation
described in the previous section.

Let us consider a large time $u$, and the corresponding
spin configuration $S_i(u)$.
Our previous observations suggest that
the value of $q_{EA}$ can be used to decompose the spin configuration
in a ``fast part'' and a ``slow part'' according to
\begin{equation}
S_i(u)= [S_i(u)-m_i(u)]+m_i(u)
\label{dec}
\end{equation}
where the slow variable $m_i(u)$ can be estimated immediately from the 
running average 
\begin{equation}
m_i(u)\simeq \frac{1}{\Delta t}\int_u^{u+\Delta t}dw\; S_i(w) \;\;\;
{\rm with}\;\;\; C(u+\Delta t,u)=q_{EA}.
\label{running}
\end{equation}
We will see later how to improve on this estimate. 
By (\ref{dec}) and (\ref{running}) we obtain correctly
the corresponding decomposition of the correlation function
in two time domains
\begin{equation}
\langle [S_i(t)-m_i(t)][S_i(u)-m_i(u)]\rangle =C_{st}(t-u)
\end{equation}
while
\begin{equation}
\langle m_i(t) m_i(u)\rangle =C_{ag}(u/t).
\end{equation}
any other decomposition, obtained averaging the spins over times such
that the value of $C(u,v)$ is different from $q_{EA}$,  would mix
$C_{st}$ and $C_{ag}$.

Through this decomposition one can define a dynamical notion of
``quasi-state'' in which the system (almost) equilibrates before
relaxing further. The quasi equilibrium hypothesis can be formulated
considering  the probability distribution of finding the system in
a given configuration of the slow and the fast variables at time $t$
induced by the
thermal noise and the flat distribution of initial
conditions
\begin{equation}
P_t(\{S_i\},\{m_i\}) =\langle \prod_i \delta(S_i(t)-S_i)\delta(m_i(t)-m_i)
\rangle_{{thermal\;\;
noise\atop initial\;\; conditions}}.
\end{equation}
which can be written as:
\begin{equation}
P_t(\{S_i\},\{m_i\}) = P_t(\{S_i\}|\{m_i\}) ) P_t(\{m_i\}) ).
\end{equation}
All the known properties of  the dynamical solution, and in particular
the short time response to external perturbations, are consistent with
the       proposition   that         the   conditional     probability
$P_t\left(\{S_i\}|\{m_i\}\right)$ becomes  independent   of  time, and
takes asymptotically the form of a restricted Boltzmann
measure:\footnote{To be more precise we should define the measure in
such a way that each $S_i$ has average $m_i$. One can check with a
detailed calculation that this condition is automatically fulfilled by
the measure (\ref{boltzmann}).}
\begin{equation}
P\left(\{S_i\}|\{m_i\}\right)=
\frac{\e^{-\beta H(\{S_i\})}\delta
\left(\sum_i S_i m_i-Nq_{EA}\right)}{\sum_{\{S_i\}}\e^{-\beta
H(\{S_i\})}\delta
\left(\sum_i S_i m_i-Nq_{EA}\right)}.
\label{boltzmann}
\end{equation}
We stress that the same property manifestly would not hold had we
chosen time scales such that $C(t,u)<q_{EA}$ in the average
(\ref{running}).  This proposition is implicit in the definition of
the aging regime, where there is no ambiguity in the definition of
$q_{EA}$. The measure should be restricted to the transverse
configuration space projecting out those directions along which the
system evolves ({\it transverse} quasi-states) where the energy
landscape is flat or has negative eigenvalues.  In these conditions,
the free-energy of the quasistate (see below), as well as the value of
$q_{EA}$ entering in (\ref{boltzmann}) are close to their asymptotic
values, but still depend on time.  We can safely assume that in the
asymptotic regime the number of negative directions become vanishingly
small.  Notice that, if we take two macroscopically different sets of
slow variables $\{m_i\}$ and $\{m_i'\}$ then, by construction, the
corresponding conditional probabilities
$P\left(\{S_i\}|\{m_i\}\right)$ and $P\left(\{S_i\}|\{m_i'\}\right)$
are mutually orthogonal. This can be easily understood from the fact
that the mutual overlap among a configuration with non-zero weight in
the first distribution and a configuration with non-zero weight in the
second one is almost surely smaller then $q_{EA}$.  It is therefore
convenient to think of a discretized $m_i$-sphere such that the
centers of the neighboring cells correspond to disjoint
quasistates. In general, we expect different quasistates to define
disjoint regions in configuration space and that the union of all such
regions define a partition of all relevant configurations. We will use
$\alpha$ as the index of the quasistate which of course will change
with the slow time. In (\ref{running}) therefore $m_i(u)$ should rather read
$m_i^\alpha$ with $\alpha$ a function of $u$. The finiteness of
$\Delta t$ limits the accuracy of the running average estimate.
 To derive a better estimate of $m_i^\alpha$ we could
clone the trajectory from time $u$ on and take a weighted average along
all trajectories.

It is useful to define thermodynamic quantities such as the
dynamical free-energy:
\begin{equation}
{\cal F}_t=\sum_\alpha \;  P_t (\{m_i^\alpha\}) \left[ F(\{m_i^\alpha\})+T
\log \left( P_t (\{m_i^\alpha\})\right)\right]
\end{equation}
where $T$ is the temperature of the thermal bath, and
\begin{equation}
F(\{m_i^\alpha\})=
\int \prod_i d S_i \;  P(\{S_i\}|\{m_i^\alpha\}) \left[ H(\{S_i\})+T
\log \left( P(\{S_i\}|\{m_i^\alpha\})\right)\right]
\end{equation}
is the free-energy of the (transverse) $\alpha$ quasi-state.

Observe that ${\cal F}_t$ includes
the average free-energy of the quasi-states
\begin{equation}
F_t=\sum_\alpha \;  P_t (\{m_i^\alpha\})  F(\{m_i^\alpha\})
\end{equation}
and  a slow entropy term
\begin{equation}
{\cal S}_t=-\sum_\alpha \;  P_t (\{m_i^\alpha\}) \log \left( P_t
 (\{m_i^\alpha\})\right).
\label{res}
\end{equation}

An explicit computation shows that, due to the disjointness property
of the quasistates, the sum $S_t+{\cal S}_t$ coincides with the
entropy of the distribution $P_t(\{S_i\})$.

 We can identify
${\cal F}_t$ with the dynamical free-energy $\int \prod_i d S_i
P_t(\{S_i\})[H(\{S_i\})+T\log P_t(\{S_i\})]$.

This last quantity is known to decrease in any process verifying
detailed balance. In our case, due to the white average over the
initial conditions we expect in addition both $F_t$ and  ${\cal S}_t$
to decrease with time.

For typical trajectories extensive quantities are self-averaging and
therefore the free-energy $F_t$ is a well defined function. The asymptotic
value, $F_\infty$, is the
free-energy of the equilibrium state for
SK-like systems, and the free-energy of the threshold TAP solutions for the
p-spin class.

The role of ${\cal S}_t$ in the dynamical relaxation is not
immediate. By construction, it does not say anything about the number
of quasi-states accessible starting from a generic point of a trajectory at
time $t$.
In fact we expect that the support of $P_t (\{m_i\})$ decomposes in
non-overlapping, mutually inaccessible, regions of phase space that
become more and more isolated as time advances.

By inverting the relation $F_t$ among free-energy and time, 
we can define ${\cal S}(F)={\cal S}_{t(F)}$ and derive that ${\cal S}
(F_\infty)$ is the extensive part of the configurational entropy
of ground states in SK-like models (zero in this case) and
the configurational entropy of the threshold states in the p-spin. The {\it
dynamical} entropy that we defined weights different regions of phase space
according to their basins of attraction. We are here using the observation
that being the threshold states identical in all their properties they must
also have the same basin of attraction. Furthermore with respect to those
states which appear in exponentially smaller numbers there is the
additional observation that their basin of attraction cannot be so much
larger as to compensate for their smaller multiplicity.

For large time, the support of
$P_t(\{m_i\})$ will be in regions of small TAP gradient. 
We can calculate ${\cal S}(F)$:
\begin{equation}
{\cal S}_t= {\cal S}(F_t)=\frac{\partial {\cal S}(F)}{\partial F}|_{F_\infty}
(F_t-F_\infty)+ {\cal S}_\infty
\label{bound}
\end{equation}

In our scenario ${\cal S}(F_t)$ measures the multiplicity of
quasistates at $t$ which as before will have equal basin of
attraction. A detailed calculation in the p-spin model is developed in
the Appendix. It shows that if we compute the multiplicity of minima
of the modulus of the gradient of the TAP free energy then their
number and derivative are continuous across the threshold value.
Therefore we propose to identify {\it the quasistates with the minima
of the gradient of the TAP free energy}. For SK-like models we
conjecture instead ${\cal S}(F)$ to be equal to the number of stable
TAP excites states at level $F$.

With this identification we now write $\frac{\partial {\cal S}(F)}{\partial
F}|_{F_\infty}$ equal to $\beta x$. It is one of those remarkable
coincidences, referred to in the introduction, that the same value of
$x$ appears in the anomalous FDR. We will interpret this coincidence
in the following sections. 

This identification of the quasistates is crucial.  An explicit
calculation of the dynamical entropy could check its validity but
unfortunately with our present techniques such calculation is not
feasible.

A strong hint in favor of it comes however from the study of the
equilibrium dynamics at temperatures slightly larger then the
dynamical transition temperature $T_d$. For $T-T_d<<T_d$ there is a
similar separation of 2 time scales controlled by the ratio
$\frac{T-T_d}{T_d}$. This allows the definition of dynamical
quasistates along lines similar to the ones followed in the
non-equilibrium case.  The free-energy obtained by considering the
collection of the quasistates of appropriate energy, should be coherent
with the direct computation from the partition function.  By an
explicit computation that we sketch in the appendix, we verified that,
up to second order corrections in $T-T_d$, the thermodynamic entropy
coincides with the TAP internal entropy plus the configurational
entropy of the saddles.

\section{Regression of Fluctuations and Onsager postulate}

In order to discuss the behavior of the response function we consider
the set-up of ``thermoremanent magnetization'' (TRM) experiment
\cite{aging}. The system is allowed to age in a small field $h$ acting
from time $t_w$ to time $u$ such that $C(u,t_w)\to 0$.  At later times
$t>u$ one detects the magnetization $M(t)= \frac 1 N \sum_{i=1}^N
S_i(t)=h \chi(t,u)$.
\footnote{It should be kept in mind that $M(t)$ denotes the
magnetization at time $t$ but can depend on both $t$ and $u$.} Our set-up
differs slightly from the one considered usually
in the literature, in which
the field acts directly from the quenching time. We switch the field on
at time $t_w$ because we find conceptually
clearer to discuss the behavior of the magnetization starting from
a situation
where the system is already in the asymptotic regime.
We notice that
the response to any arbitrarily varying field $h(t)$ can be expressed as a
linear superposition of TRM magnetizations.

  In order to discuss the decay of $M(t)$ we will show that a
generalization of the Onsager postulate of normal regression of
fluctuations applies to the dynamical off-equilibrium process
\cite{onsager}. The principle, originally stated for equilibrium
systems, concerns the behavior of {\it macroscopic quantities} and
states that in the linear response regime one can not distinguish the
regression of a spontaneous fluctuation of a certain quantity from the
regression from the same value when imposed through a constraint on the
equilibrium measure.
Onsager's postulate means that for a large system,
a spontaneous fluctuation must have the
characteristic of
the {\it most probable} fluctuations and therefore  correspond to
constrained minimization of the free-energy. This is equivalent to
free-energy minimization in a conjugated field, thus leading
to an immediate
derivation of
the fluctuation-dissipation theorem.

More in detail the
argument can be phrased as follows \cite{onsager}.  Consider a
thermodynamic system at equilibrium and a given macroscopic
(extensive) quantity $\alpha$ which takes the value zero at
equilibrium. Be $\gamma$ the corresponding conjugate intensive variable.
Suppose that at time zero the quantity $\alpha$ has a small
but extensive spontaneous fluctuation $\alpha_0$.  This will occur
with exponentially small probability, but when it occurs the
subsequent evolution of $\alpha(t)$ will be independent of the thermal
noise, i.e. $\alpha(t)=E(\alpha(t)|\alpha_0)$, where we denoted by
$E(\cdot|\alpha_0)$ the conditional expectation over the trajectories
for fixed $\alpha_0$ at time zero.  As $\alpha_0$ is small, we can write
\begin{equation}
\alpha(t)=E(\alpha(t)|\alpha_0)=A(t)\alpha_0
\label{uno}
\end{equation}
Denoting by
$E_{\alpha_0}(\cdot)$ the average over the distribution of $\alpha_0$,
and $C_\alpha(t)=E(\alpha(t)\alpha(0))$ the correlation function,
it follows that $A(t)=C_\alpha(t)/ E_{\alpha_0}(\alpha_0^2)$.
Notice that the typical values of $\alpha_0$ entering in
the correlation function are of the order $\sqrt{N}$, while in the
relation (\ref{uno}) we consider values of order $N$. The validity
of the analysis above relies on the smoothness of the probability
distribution of $\alpha_0$ in the crossover region,
assumption which is at
the heart of linear response theory.

As one is conditioning
(\ref{uno}) by the value of $\alpha_0$ only, then the overwhelming
majority of the configurations ${\cal C}$ giving rise to the
fluctuation are the ones ``typical'' of the restricted canonical
distribution
\begin{equation}
\frac{\e^{-\beta H({\cal C})}\delta(\alpha({\cal C})-\alpha_0)}
{\int d{\cal C}\e^{-\beta H({\cal C})}\delta(\alpha({\cal
C})-\alpha_0)}.
\label{m1}
\end{equation}
which is equivalent to
\begin{equation}
\frac{\e^{-\beta (H({\cal C})-\gamma\alpha({\cal C}))}}
{\int d{\cal C}\e^{-\beta (H({\cal C})-\gamma\alpha({\cal C}))}}
\label{m2}
\end{equation}
in which $\gamma$ is fixed by: $\langle
\alpha\rangle_\gamma=\alpha_0$. Since to the linear order in $\gamma$
we have $\langle \alpha\rangle_\gamma=\beta \gamma \langle
\alpha^2\rangle_{\gamma=0}$, it follows that the relaxation of
$\alpha(t)$ induced by the field is given by:
\begin{equation}
\alpha(t)=\beta\gamma C_\alpha (t).
\label{fdt}
\end{equation}
which is the fluctuation-dissipation theorem in
its integral form.

Here we would like to show how a generalized form of the regression
principle holds in aging dynamics where the time scale separation
suggests that, besides
the fluctuations of the instantaneous magnetization $M(u)$ one should also
consider possible fluctuations of the running global magnetization $m(u)$,
defined as
\begin{equation}
m(t)= \frac 1 N \sum_{i=1}^N m_i(t).
\label{run}
\end{equation}

We consider the conditional expectation
value of the magnetization at time $t$ given small values of the
instantaneous and running magnetizations $M(u)$ and $m(u)$: $E\left(
M(t) | M(u),m(u)\right)$. This can again be expanded to the
first order:
\begin{equation}
E\left( M(t) | M(u),m(u)\right)=A(t,u)[M(u)-m(u)]+B(t,u) m(u).
\end{equation}
and the functions $A$ and $B$ can be fixed by a hypothesis of
continuity, leading to

\begin{equation}
E\left( M(t) | M(u),m(u)\right)=\left[
\frac{C_{st}(t-u)}{1-q_{EA}}[M(u)-m(u)]+
\frac{C_{ag}(u/t)}{q_{EA}} m(u)\right].
\label{cond}
\end{equation}
where we have used $\langle (M(u)-m(u))^2\rangle=\frac{1-q_{EA}}{N}$
and $\langle m(u)^2\rangle=\frac{q_{EA}}{N}$.

Onsager's argument demonstrates two things:
\begin{enumerate}
\item
that the decay of a spontaneous fluctuation with time is
governed by the correlation function.
\item
that a fluctuation induced by a conjugate field will decay as a
spontaneous fluctuation if the probability distribution
defining the
state of the system immediately after the induced field is turned off is
equal to the unperturbed probability distribution projected on the
hypersurface defined by the equations:
\begin{equation}
\frac{1}{N}\sum_{I=1}^N m_i(u)=m(u); \;\;\;\; \frac{1}{N}\sum_{I=1}^N
S_i(u)=M(u),
\end{equation}
where now, $m(u)$ represents the value of an average as
(\ref{running}) for times immediately {\it after} the field is turned off.
\end{enumerate}
This second condition is consistent with our scenario of quasi-equilibrium
in the dynamical relaxation process. In the next sections we
will deal with the problem of computing the slow part of the
magnetization induced by a field. We will first
discuss the case of SK-like models whose OTOs during the dynamical
relaxation tend to the ground state values. Then  we will
discuss those systems where the asymptotic state is
different from the ground state. In this case, the argument is
further complicated by the extensive multiplicity of the threshold
states.

\section{The case of SK-like models.}

We first recall that the equilibrium analysis of these models
\cite{MPV} in the ``one step replica symmetry breaking'' approximation
determines the multiplicity of states at low free-energy $F$ \cite{mpv}
as
\begin{equation}
{\cal N}(F) dF =\e^{\beta x (F- F_{GS})} dF
\label{23}
\end{equation}
where $F_{GS}$ is the ground state free-energy and $x$ is the Parisi
parameter in this approximation.

We then quote from the dynamical solution the expression of the
magnetization in the TRM experiment described in the previous section:
\begin{equation}
M(t)=C_{st}(t-u)\beta h+C_{ag}(u/t)\beta h x.
\end{equation}

The comparison of this with equation (\ref{cond}) tells us the following
remarkable fact: the action of an external field $h$ from time $t_w$ to $u$
produces at $u$ a state of the system (in the sense of a measure in the
microscopic variables) which is identical to the one we can obtain through
infinite realizations of the thermal noise and selection of those
trajectories with
\begin{eqnarray}
& & M(u)-m(u) =\beta h (1-q_{EA})\nonumber\\
& & m(u) = \beta h x q_{EA}.
\label{small}
\end{eqnarray}

Therefore, thanks to the use of Onsager's postulate it is enough to
calculate the response at time $u$ immediately after the magnetic
field has been turned off. We have assumed $t_w$ and $u$ sufficiently
large so that the system is in a quasi-state with free-energy $F$
slightly larger then the one of the ground state. Eqs. (\ref{small})
separate the response to the magnetic field in two components: a)
inside the same quasi-state the more probable configurations will
change and b) the quasi-state will change. The response a) is the
equilibrium intrastate response and is trivial. To isolate b) we
imagine turning off the magnetic field at time $u$ and then waiting a
finite time $\Delta t$ such that $C_{st}(\Delta t)$ is $q_{EA}$ while
still $\Delta t/u$ is zero. Then we know that the system has gone from
one quasi-state at time $t_w$ to another at time $u+\Delta t$
both defined with zero
magnetic field. The distribution of (zero magnetic field) quasi-states
with this free-energy is given by (\ref{23}).  Each of them may have a
magnetization, uncorrelated from the free-energy and with variance
$\langle m^2\rangle = q_{EA}/N$.  The typical number of quasi-states
with free-energy density $F$ and magnetization $m$ is therefore given
by
\begin{equation}
{\cal N}(F,m)  =\e^{\beta x (F- F_{GS})} \e^{-N\frac{m^2}{2q_{EA}}}
\end{equation}
implying that
\begin{equation}
{\cal S}(F,m) = {\beta x (F- F_{GS})}{-\frac{m^2 N}{2q_{EA}}}\geq 0
\end{equation}
We first note that if we send $u$ to infinity before sending $h$ to
zero, i.e. we consider fields such that the induced magnetization $m$
verifies $\beta x (F_u- F_{GS})<< {m^2 N}/{2q_{EA}}$ we can derive the
result (\ref{small}) in a quite straightforward way. In fact, we obtain
that a non zero magnetization has to be compensated by an increase of
free-energy so as to keep the configurational entropy ${\cal S}(F,m)$ non
negative
\begin{equation}
F-F_{GS}={\frac{m^2 N}{2q_{EA}\beta x }}.
\end{equation}
Along this line in the $F,m$ plane the state with
lowest total free-energy $F-hmN$ has
\begin{equation}
F=F_{GS}+{\frac{\beta N h^2 q_{EA} x}{2 }}.
\end{equation}
implying $m=\beta x h q_{EA}$.

The interpretation of this result is particularly
illuminating. Turning on the magnetic field is a way of making
energy available to the system. The thermal bath would normally absorb
part of this energy. However this is possible
only if the entropy of the system decreases in the process. This
cannot happen here as by hypothesis the available entropy
is much smaller than the one required to increase $m$.
We conclude that the equilibration must occur only
between the magnetic free-energy $hmN$ and the unperturbed, zero magnetic
field  $F$.

With this argument in mind we can now understand the limit more
relevant to the dynamical approach. In this case $F(u)- F_{GS}$ is
extensive and large with respect to the potential energy introduced by
the external magnetic field.  In this situation there is,
formally, enough entropy to allow the magnetization to reach the value
of equilibrium with the thermal bath $m=\beta h q_{EA}$. However, with
the same token one would argue that the thermal bath could have
absorbed that entropy to decrease the  spin-spin
interaction energy. We know that this is not the case, or rather that
entropy/heat is absorbed at a certain rate basically determined by the
barriers. The external force is uncorrelated with the direction of
relaxation of the system, and therefore it is reasonable to assume that the
turning on of the magnetic field {\it will not modify the rate of entropy
decrease (heat transfer to the thermal bath)}. We conclude as before that
the equilibration must occur
between the magnetic free-energy $hmN$ and $F$. In formulas if we call
$F^h(u), F(u)$ the free-energy (associated with the inter spins couplings)
that the system would reach in the presence of the magnetic field or in its
absence at time u, then:
\begin{equation}
{\cal S}(F^h(u),m(u)) ={ \cal S}(F(u))
\end{equation}
so that:
\begin{equation}
{\beta x (F^h(u)- F_{GS})}{-\frac{m(u)^2 N}{2q_{EA}}}=
{\beta x (F(u)- F_{GS})}
\end{equation}

The previous argument now follows minimizing $F^h(u)-Nhm(u)$.

We remark that both
entropy reductions refer to the same degrees of freedom, and therefore
respond on the same time scale.
The result is that the thermal bath acts as if it was uncoupled while the
two forms of (free-) energy mutually
equilibrate.\footnote{This represents an instance of the
recent proposal that a system and a thermometer responding in the same time
scale will
equalize their effective temperatures \cite{teff}. In fact the inverse
temperature of the magnetic field interaction energy is
$dS/dE_h=d(-m^2/2q_{EA})/d(-mh)=m/(q_{AE}h)=\beta x$. However our entropic
interpretation suggests
that time scales will strongly depend on $\beta x$, the lower the
effective temperature, the slower the evolution of the system;
if  two different aging
systems starting with different effective temperatures and equal time scales
are put in contact, they
will quickly develop different time scales before equilibrating.} In other
words the transition time to higher free-energy states is much smaller than
the one required to go to equal or lower free-energy states.

This argument is so crucial to our picture that we feel it necessary to
try to confirm it with a detailed model of the dynamical process.

Let us imagine the dynamical trajectory from a (large) time $u$ to a time $t$
such that $C(t,u)\approx 0$. We discretize the dynamics in $k$ steps
such that $u=t_0<t_1<...<t_k=t$ such that
$C(t_{i+1},t_i)=q_{EA}-\epsilon$ with $\epsilon$ small. At time $t_i$
the system will have free-energy $F_i$ and $F_{i+1}-F_{i}$ will be
small but extensive. The model we make of the dynamical process
consists in assuming that when going from time $t_i$ to time $t_{i+1}$
the system can access different quasi-states with the lower one at free-energy
$F_{i+1}$ and the higher ones distributed exponentially
\begin{equation}
{\cal N}(F)= \e^{\lambda (F-F_{i+1})}
\end{equation}
while the probability of transition to a state with free-energy $F$ is
proportional to
\begin{equation}
\e^{-\beta F}
\end{equation}
The model is consistent for $\lambda<\beta$, (otherwise the free-energy
would grow with time) and  incorporates the following two
features:
\begin{enumerate}
\item
The decrease in extensive free-energy is deterministic
\item

If we fix the initial condition, the increase in entropy in a single
step is finite.\footnote{The entropy about which
we are talking here correspond to a the dynamical probability in which
the initial condition is fixed, and is
therefore increasing with time.}
In fact this can be calculated following the lines in
\cite{gross} for the Random Energy Model, with the result
\begin{equation}
\Delta
S=\Gamma'(1)-\frac{\Gamma'(1-\lambda/\beta)}{\Gamma(1-\lambda/\beta)}
\end{equation}
\end{enumerate}
In $k$ steps the entropy generated will be $k\Delta S$ and therefore
negligible with respect to $N m^2/(2q_{EA})$. Although non-extensive,
$k\Delta S$ can be arbitrarily large, thus explaining the divergence
of two cloned trajectories. In this model we have heavily used the
self-averaging character of the macroscopic quantities along the
trajectories. It is again clear that the only way to develop a
magnetization is by compensating it with an increase in the (zero
magnetic field) free-energy.

\section{p-spin like models}

For p-spin like models even the limit $t_w\to\infty$ before $h\to0$ is
non-trivial.  In fact it is well known that for this kind of systems
the properties of the quasi-states encountered in the dynamics are
closer and closer to these of the threshold TAP states, which have
extensive configurational entropy $S_{th}$. If this entropy would be
accessible in the dynamical process the equality (\ref{bound}) would
be valid with ${\cal S}_\infty=S_{th}$.  The condition that the total
configurational entropy at time $u$ be positive would then read
\begin{equation}
S_{th}+{\beta x (F_u- F_{th})}{-\frac{m^2 N}{2q_{EA}}}\geq 0
\end{equation}
which could be satisfied even if $F_u- F_{th}$ is small and negligible
in front of ${-\frac{m^2 N}{2q_{EA}}}$. In more physical terms we can
say that among the $\e^{S_{th}}$ states there are
$\e^{S_{th}{-\frac{m^2}{2q_{EA}}}}$ states with magnetization $m$. If
all these states were available to a single trajectory the response
 would be normal, $m=\beta h q_{EA}$. If we want the response
to be anomalous we must show that the system, while wandering in phase
space has no access to the configurational entropy.\footnote{A moment
of reflection reveals that otherwise the system, wandering in such a
large space, would pass to lower lying states and relax below $F_{th}$.}

The logarithm of the number of states in the vicinity of any given TAP
state has been computed by Cavagna, Giardina and Parisi in
ref. \cite{cgp} in the case of the $p$-spin model. Below threshold
all states  are
isolated; there are no states closer than a given distance
$q_{EA}-q_{max}$.  $q_{max}$ is a level-dependent overlap which
tends to $q_{EA}$ at threshold. Right at the threshold, the
logarithm of the number of states
 as a
function of the distance $(q_{EA}-q)$ is
\begin{equation}
N\Sigma(q)\propto N (q_{EA}-q)^5.
\end{equation}

  Let us again imagine a discretization of the dynamics in which at
each step the system can jump a distance $\delta=(q-q_{EA})$. Then
after $n$ steps the log of the number of accessible states would be at
most of the order of $n\;\delta^5$.\footnote{This estimate could
correspond to a severe double counting, as one can realize applying
the estimate to finite dimensional Brownian motion. In infinite
dimensional problems we expect it to give essentially the
correct result.But, in any case, we only need it as an upper bound
in our argument.}  On the other hand the distance traveled will be:
$\Delta=n\; \delta$ if all the steps are in the same direction and
$\Delta=\sqrt{n}\;\delta$ if the steps are uncorrelated. In both cases
it is easy to see that if we take the limit $\delta\to 0$ and
$n\to\infty$ fixing $\Delta$ we find that $log({\cal N})/N$ goes to
zero ($n\;\delta^5\to 0$).

Notice that the argument is based on the scarcity of states in the
vicinity of a given state. This should be a generic feature for p-spin like
systems  other then the $p$-spin model.

Having eliminated the configurational entropy from the balance,
the
argument proceeds as in the case of
SK-like models.

Let us conclude by pointing out that threshold states with large
magnetization (of order $\beta h q_{EA}$) do exist, but are non-critical
in the presence of the field. Therefore with probability one such
states would be isolated and unreachable.

\section{Summary and conclusions}

 The main point of our analysis has been to give an explanation of the
anomalous response function.  We have found the physical origin of the
equality between the FDR and the growth rate of the configurational
entropy close to the asymptotic state. The value of the anomalous
response can be traced to the lack of available entropy when the
system is close to the low lying states.  Our interpretation clarifies
the relation among equilibrium properties and off-equilibrium
dynamics.  For p-spin-like systems we have argued that the
extensive configurational entropy of the threshold states does not
play any thermodynamical role. We have seen that the classical
Onsager's argument on the equivalence between the regression of a
spontaneous, noise-caused, fluctuation of the magnetization and the
one induced by an external field can be generalized to aging
systems.

Our analysis can be summarized by saying that in aging systems
the rate of entropy decrease
is a function of age and does not change due to small forces. Thus the
balance is  always between the value of the
unperturbed free-energy and the one of the perturbation, without
taking into account the thermal bath. We expect this conclusion to
hold also in short range systems with aging.

In spin glass materials, one time observables (OTO)
equilibrate, and the picture we have developed relates to the
structure of configuration space close to the ground state. In aging
experiment of structural glasses on the other hand, OTOs are far from
their asymptotic values. Still, one can observe quasi-scaling aging
dynamics on two time observables. The structure of the phase space
visited on this time scale can not be related to ``true'' asymptotic
properties of the system. We would like to speculate that even here
the fluctuation-dissipation ratio, which could be a slowly varying
function of time, is related to the derivative
of available phase space with free energy also varying along the dynamical
path. This could be true even if the system would
eventually reach equilibrium on a different time scale where FDT is
asymptotically obeyed.

Finally the case of multiple time sectors or multiple replica symmetry
breakings
will need trivial modifications.

\section*{Acknowledgments}

We thank M. Mezard, R. Monasson, G. Parisi, and L. Peliti
for important discussions at the early stages of this work.

\section{Appendix}

The aim of this appendix is two fold. We first show that in the p-spin
model the derivative of the configurational entropy of the saddles is
continuous at the threshold.  Then, we prove that above $T_d$ the
paramagnetic state can be seen, to the first order in $T-T_d$ as a
collection of quasi-states identifiable with the points of least
gradient of the TAP free-energy.

Let us start from the expression of the TAP free-energy for the p-spin
model \cite{KPV}
\begin{equation}
F_{TAP}[m_i=\sqrt{q}S_i]={E_0}\,{q^{{\frac{p}{2}}}} - 
  {\frac{\beta\,\left( 1 - 
        p\,\left( 1 - q \right) \,{q^{-1 + p}} - 
        {q^p} \right) }{4}} - 
  {\frac{\log (1 - q)}{2\,\beta}}
\label{fTAP}
\end{equation}
where $E_0$ is the angular part of the energy as a function of the
angular variables $S_i$. It is well known that while one can find
stationary points of the angular part for all the values of $E_0$ in
the range $|E_0|>-E_{GS}$.  Conversely, at finite temperature one finds
solutions for the radial parts only in the range
$-E_{th}>|E_0|>-E_{GS}$. The overwhelming majority of these solution
are free-energy minima. 

The stationary points of the angular part for $-E_{th}<|E_0|$ turn out
to be saddles, with a number of unstable directions which depends on
$E_0$. The number of stationary points as a function of $E_0$ is given
by \cite{criso}
\begin{equation}
\Sigma(E_0)=\frac{1}{2}\left[{\frac{2 - p}{p}} - 
     {\frac{2}{{p^2}\,{z^2}}} + 
     {\frac{\left( -1 + p \right) \,{z^2}}{2}} - 
     \log ({\frac{p\,{z^2}}{2}})\right]
\label{sig}  
\end{equation}
where $z$ is an auxiliary variable given by
\begin{equation}
-{\frac{{E_0}}{-1 + p}} - 
  {\frac{{\sqrt{{{{E_0}}^2} - 
         {{{E_{th}}}^2}}}}{-1 + p}}
\end{equation}
For the saddles $E_0>E_{th}=-\left( {\sqrt{2}}\, {\sqrt{{\frac{-1 +
p}{p}}}} \right)$ the formula become complex. This is due to the fact
that the Hessian which appears in the computation \cite{criso} has negative
eigenvalues and one has to compute the absolute value of its
determinant. As suggested in \cite{cgp} this can be done just taking
the real part of expression (\ref{sig}), which gives the parabolic shape
\begin{equation}
\Sigma(|E_0|<-E_{th})=- E_0^2\,
        \frac{( p-2)}{2(p-1)}  + 
        \,
       \frac{1}{2} \log (
           p-1)
\label{sopra}
\end{equation}
An explicit computation using this formula shows that the $E_0$-derivative
of (\ref{sig}) and (\ref{sopra}) is continuous at the threshold
energy.

Let us now pass to our second task.  We would like to identify the
quasistates close to threshold as points of minima of TAP gradient.
Unfortunately we were not able to prove this directly in the aging
regime at low temperature, for we do not know how to compute the
dynamical entropy. We start then from the observation that for
temperatures higher, but close to $T_d$ one observes slowing down of
the dynamics with time scale separation which is less and less
ambiguous as $T\to T_d$. So we can define dynamic quasistates even
above $T_c$, where the role of a small but finite $T-T_d$ is similar
to the role of a finite $t_w$ in the low temperature dynamics.  We put
these quasistates in relation with the TAP free-energy, supposing that
they coincide with the points of least TAP gradient for fixed internal
energy equal to the paramagnetic value $-\beta/2$.  These are saddle
points of the angular part, while the radial part is an inflection point,
i.e. we fix $q$ in the value of the minimum of $d F_{TAP}/d q$.  By
explicit computation from (\ref{fTAP}) and (\ref{sopra}) we find that
the total free-energy $F_{TAP}-T\Sigma(E_0)$ is equal to the
paramagnetic free-energy $-\beta/4$ up to terms which are of the
second order in $T-T_d$. For instance for $p=3$ one finds that
$F_{TAP}-T\Sigma(E_0)=-\beta/4-8\sqrt{2/3}(T-T_d)^2$.

\end{document}